\documentclass[aps,prd,preprint]{revtex4}
\usepackage{epsfig}
\usepackage{latexsym,amsmath,amssymb,amstext}
\usepackage{float}
\usepackage{graphicx,xcolor}

\newcommand{\be}{\begin{equation}}
\newcommand{\ee}{\end{equation}}
\newcommand{\bea}{\begin{eqnarray}}
\newcommand{\eea}{\end{eqnarray}}

\newcommand\bef{\begin{figure}}
\newcommand\eef[1]{\label{fg:#1}\end{figure}}
\newcommand\beq{\begin{equation}}
\newcommand\eeq[1]{\label{#1}\end{equation}}
\newcommand\beqa{\begin{eqnarray}}
\newcommand\eeqa[1]{\label{#1}\end{eqnarray}}
\newcommand\bet{\begin{table}}
\newcommand\eet[1]{\label{tb:#1}\end{table}}

\newcommand\fgn[1]{Figure \ref{fg:#1}}
\newcommand\eqn[1]{Eq.\ (\ref{#1})}

\newcommand{\fm}{\text{fermi}}

\begin{document}
\title{Scale invariant behavior in a large N matrix model}
\author{Rajamani\ \surname{Narayanan}}
\email{narayanr@fiu.edu}
\affiliation{Department of Physics, Florida International University, Miami, FL 33199.}
\author{Herbert\ \surname{Neuberger}}
\email{neuberg@physics.rutgers.edu}
\affiliation{Department of Physics and Astronomy, Rutgers, The State University of New Jersey, Piscataway, NJ 08854.}

\begin{abstract}
Eigenvalue distributions of properly regularized Wilson loop operators are used to study the transition from ultra-violet
(UV) behavior to infra-red (IR) behavior in gauge theories coupled to matter that potentially have
an IR fixed point (FP). We numerically demonstrate emergence of scale invariance in a matrix model that describes $SU(N)$
gauge theory coupled to two flavors of massless adjoint fermions in the large $N$ limit. The eigenvalue distribution of Wilson loops of varying sizes cannot be described by a universal lattice beta-function connecting the UV to the IR.  
\end{abstract}

\date{\today}

\maketitle

The Lagrangian of four dimensional gauge theories coupled to massless fermions has no dimensionful parameter.
After applying a proper regulator to cure ultra-violet (UV) divergences, the physics in the infra-red (IR) limit is
expected to fall into two classes: One provides a quantum description of relativistic particles where the scale
invariance of the classical Lagrangian is broken; the second describes a conformal theory with no particle content.
QCD, as observed in nature, belongs to the first class.
In both classes, the physics at asymptotically short distance scales is described by perturbation theory in terms of a redundant set
of local degrees of freedom. The physics at large distances, where the two classes are different, can be  easily characterized in terms of non-local observables.

Models in the first class which are ``borderline'' in terms of their proximity to the second class are central to an activity 
evaluating certain proposals for beyond the standard model physics~\cite{Sannino:2009za,Lucini:2015noa}. The present work does not consider this issue. For us, the mere existence of an interacting conformal gauge theory in four dimensions which seems to require no fine tuning on the lattice to boot, makes the study of class two models deserving 
of attention on its own. 

Let $a$ denote an UV regulator with dimensions of length and let $g$ be the bare coupling introduced into the Lagrangian.
This enables one to perform a calculation that produces a finite result at a fixed $a$ and $g$. We define a  dimensionless physical coupling, $g_R(\ell;a,g)$, that depends on the physical scale $\ell$ and obeys 
a Renormalization Group (RG) equation:
\be
\left [ a\frac{\partial}{\partial a} +\beta(g)\frac{\partial} {\partial g}\right ]g_R(\ell; a,g)=0;
\ \ \ \ \ \beta(0)=0; \label{rgeqn}
\ee
for every choice of $\ell$. 
The content of the equation is that $g$ depends on $a$ via
\be
\frac{d g}{d\ln a} + \beta(g) = 0;\label{runcoup}
\ee
and we can write our observable as $g_R(\ell;g(a))$. Noting that 
\be
\left[ \frac{\partial}{\partial \ln \ell} + \frac{\partial}{\partial \ln a}\right] g_R(\ell;g(a))=0,
\ee
we change variable from $g$ to $g_R(\ell;g(a))$ for a fixed $\ell$ 
and obtain the renormalized version of \eqn{runcoup} as
\be
\frac{\partial g_R(\ell)}{\partial \ln \ell} = \beta_R(g_R(\ell));\ \ \ \ \ \beta_R(0)=0; 
\ee
where the reference to the the UV regulator, $a$, naturally drops out. 
In perturbation theory,
 the first
two coefficients in the expansion of the renormalized beta function,
\be
\beta_R(g_R(\ell)) = \beta_0 g_R^3(\ell) + \beta_1 g_R^5(\ell) + \cdots,\label{perbeta}
\ee
are the same as that of the unrenormalized beta function $\beta(g)$. It is tempting to define a new renormalized coupling where
the associated beta function is exactly given by the first two terms in \eqn{perbeta}. This new renormalized coupling will admit a Taylor expansion in terms of the old coupling around the joint value 0. There is no guarantee that this new renormalized coupling and the bare coupling $g$ are related by a map that is 
well behaved at all points away from the origin.

If we assume that $\beta_R(g_R(\ell))$ remains positive for all $\ell > 0$, it follows from the RG equation for $g_R(\ell)$ that it grows monotonically without bound as $\ell$ increases, starting from $g_R(0_+)=0_+$ . There is only one length scale in such a theory: A scale, $\ell_0$,  beyond which higher order terms in
\eqn{perbeta} begin to matter and the theory becomes non-perturbative. In a theory like QCD, $\ell_0$ is close to 1 \fm\  and long
distance physics can be numerically studied with small systematic errors by simulating QCD in a periodic box with a linear extent of
few \fm. 

Now consider theories where $\beta_R(g^*)=0$ and becomes negative for $g_R > g^* >0$.  The renormalization coupling constant will not grow
without bound; instead, it will start from $g_R(0_+)=0_+$ and reach $g_R(\infty) = g^*$. In such a theory, we expect that two physically relevant length scales $\ell_s \ll \ell_l$ exist, 
such that $g_R(\ell)$ will be governed by perturbation theory for $\ell < \ell_s$ and become scale invariant for $\ell > \ell_l$.
In order to see the onset of scale invariance, a numerical study should be performed in a periodic box with a linear extent significantly larger than $\ell_l$ and this will be practically  impossible if also $\ell_s \ll \ell_l$ holds. Two scale problems are notoriously difficult to control by numerical methods. They are also a challenge to analytical methods.

We will consider $SU(N)$ gauge theory with $n_M$ flavors of massless Majorana fermions in the
adjoint representation in this paper. For this theory~\cite{Caswell:1974gg},
\be
\beta_0 = \frac{ N \left( 11 - 2 n_M\right)}{48 \pi^2};\ \ \ \ \
\beta_1 = \frac{ N^2 \left( 17 - 8 n_M\right)}{384 \pi^4}.\label{beta01}
\ee
If $\frac{17}{8} < n_M < \frac{11}{2}$, we have $\beta_0 > 0$ and $\beta_1 < 0$ and
\be
\frac{5}{7} \ge \frac{{g^*_2}^2N}{8\pi^2} \ge  \frac{1}{23}  \ \ \ \ \ {\rm for}\ \ \  3 \le n_M \le 5,
\ee
as given by the first two terms in \eqn{perbeta}. It is unlikely, this estimate of $g^*$ is correct even for $n_M=5$.

The lattice action for massless fermions coupled to gauge fields has one coupling, $b=1/\lambda$, where $\lambda=g^2N$ is the 't Hooft coupling. 
The aim of a lattice study of a theory that possibly exhibits scale invariance in the IR is not necessarily to identify a RG trajectory along which to carry out simulations, 
since a trajectory connecting $\ell_s$ to $\ell_l$ might require a path visiting actions containing a large number of terms with many couplings that need to be precomputed. 
Keeping only one coupling $b$, all we can say with some certainty is that at large $b$ its variation is tangential to this RG
trajectory. Once the regime of large $b$ is left we likely wander off quite far from any RG trajectory corresponding to some decent RG transformation. A ``decent'' RG trajectory is defined by a RG transformation in the space of actions which respects all important symmetries and is local, in the sense that the new fields are local functionals of the old fields. Also, it is required to check that the space of couplings one needs to keep track of is of a low dimension to good numerical accuracy (that is most of the couplings of the infinite space of actions can be ignored). 

The continuum situation may be taken to simply indicate that for
more or less all $b$'s the single coupling system is critical, and its
large distance asymptotic behavior is scale and conformal invariant. We don't ever sit at, or are close to, the IR fixed point of some RG map: only the large distance behavior is governed by some IR FP. This is the generic case for a critical system.
For a given $b$ there will two scales, $L_s(b) << L_l(b)$ in lattice units.
For $L < L_s(b)$ scaling is violated as in QCD and for $L>L_l(b)$ 
scaling is restored but the dimensions take ``anomalous'' values. As we vary $b$ in some range, $b \in (b_0,b_\infty)$, the  $L_s(b) << L_l(b)$ vary, but outside this range 
their very definitions are in jeopardy and may become inconsistent. 
For intermediate scales $L$, $L_s(b) < L < L_l (b)$, the most plausible behavior is one strongly dependent on the form of the lattice action.
There, lattice observables cannot be described by a continuum action on
some RG trajectory. In short, the UV - IR interpolation offered by a single coupling lattice action does not necessarily admit a useful continuum description throughout, but only at sufficiently short and sufficiently long scales. 

Large $N$ reduction enables one to reduce the lattice action to a matrix model~\cite{Eguchi:1982nm,  GonzalezArroyo:1982hz}. In some cases, the matrix model reproduces
physics at infinite volume lattice theory.
 For QCD-like theories with the
number of Dirac flavors in the fundamental representation kept finite
this connection breaks down
as one approaches the continuum limit $b\to\infty$~\cite{Bhanot:1982sh}.  This breakdown
can be avoided if one uses more than a modest amount of adjoint matter~\cite{mkrtchyan:1982mk,Kovtun:2007py,Hietanen:2009ex}. 
Assuming no other problems with large $N$ reduction, this offers the opportunity to look for asymptotic scale invariance in a large $N$ matrix model. 

We will require non-local observables to properly study the IR behavior. 
A basic set of non-local operators in the continuum or on the lattice is given by Wilson Loops (WL). We proceed using  continuum language.
Classically, for $SU(N)$, these are unitary matrices associated with closed smooth spacetime curves $\mathcal C$. Quantum mechanically one needs to 
renormalize, giving the loop an effective thickness:
\be
W_f(\mathcal C,x,s)=\mathcal{P} \exp\left(i\oint^x_{x;\mathcal{C}} A^f_\mu(y,s) dy_\mu
\right)  \in SU(N)
\ee
$f$ denotes the fundamental representation of $SU(N)$ and $x$ is a marked point on $\mathcal{C}$. $s$ is a smearing parameter,  providing the thickness. 
The $N\times N$ matrix $W_f$ has operator valued  entries which obey
$[ W_f W_f^\dagger]_{ij}=\delta_{ij}\mathbb{I},\;\det(W_f)=\mathbb{I}$. 

Starting from the four dimensional quantum field, $B^f_\mu(x)$; $x\in R^4$, appearing in the path integral, the act of smearing~\cite{Narayanan:2006rf,Lohmayer:2011si}
extends
the gauge fields to the five dimesional space, $R^4\times R_+$, with the smearing parameter $s \in R_+$.
The five dimensional gauge fields, $A^f_\mu(x,s)$ are 
defined for $s\ge 0$ by
\be
F_{\mu s} = D_{\nu} F_{\mu\nu}; ~~~~~~~~~~~~~{A^f_\mu(x,s=0)=B^f_\mu(x)}.\label{smearing}
\ee
Note that  $s$ has dimensions of area.
The 5D gauge freedom is partially fixed, leaving a 4D one, by $A^f_s(x,s)=0$.
All divergences coming from coinciding spacetime 
points in products of renormalized elementary fields
are eliminated for $s>0$ by a limitation on the resolution of the observer, parametrized by $s$. In QCD, a reasonable value for $s$ for a loop
of size $\sim 1\;\fm$ is $s\sim 0.05 \ \fm^2$.

Let us focus on square Wilson loops of side $\ell$, noting that the effects of the sharp corners have also been smoothed out by the smearing.
For every such $W_f(\ell,x,s)$ we consider its set 
of eigenvalues 
\be
\Theta(\ell,s)=\{e^{i\theta_1},...,e^{i\theta_N}\},\ \ \ \ \ \sum_{i=1}^N\theta_i=0\mod 2\pi.
\ee
The label $x$ drops out since the eigenvalues do not depend on the choice of the point $x$ on $\mathcal{C}$.
$\Theta(\ell,s)$ parametrizes the
gauge invariant content of $W_f(\ell,x,s)$.
Using the eigenvalues, $\Theta(\ell,s)$, we can define vacuum averages which provide the complete information about the  $\Theta(\ell,s)$ in terms of $n$-angle densities $p(\Theta;\ell,s)$ normalized to unity. $p$ contains a periodic $\delta$-function representing the $\det=1$ constraint. 
Instead of working with two dimensional parameters $(\ell,s)$ 
we change the variables to $(\ell, f=\frac{s}{\ell^2})$.  
We focus on the simplest marginal of $p$, the normalized single-angle eigenvalue distribution 
\be \rho(\theta ,f; \ell) = \frac{1}{N} \sum_{i=1}^{N} \left[  \prod_{j=1}^N \int_0^{2\pi}  d\theta_j \right] p(\Theta,s=f\ell^2;\ell) \delta(\theta-\theta_i);
\ \ \ \  \int_0^{2\pi} d\theta \rho(\theta ; \ell,f) =1.
\label{single-angle}
\ee
$\rho$ carries no information about correlations between the various $\theta_i$.
The question 
of interest is how the dimensionless $\rho(\theta,f; \ell)$  
depends on $\ell$ for a fixed $f$. 

In the case of pure gauge theory,  $\rho(\theta,f; l)$ provides an acceptable definition 
of a $\theta$ independent $\beta$-function because the $\theta$-dependence admits an accurate parametrization by one variable only. The observable, $\rho(\theta,f;\ell)$, has proved to be quite useful in understanding the transition from weak coupling to strong coupling in the
large $N$ limit of QCD~\cite{Narayanan:2006rf,Narayanan:2007dv,Lohmayer:2011nq}. In the 't Hooft $N=\infty$ limit oscillations in $\theta$ disappear from $\rho(\theta,f; \ell)$ and $\rho$ is very smooth almost everywhere, but a non-analyticity in the dependence on $\ell$ appears at some $\ell_c$. 
For $\ell <\ell_c$ the support of $\rho(\theta,f; \ell)$ is restricted to an
arc $<2\pi$ symmetrically around $\theta=0$. For $\ell >\ell_c$ the support is the entire unit circle. For any finite $N >>1$ the
transition is smoothed out.  For $N>>1$ 
the $\ell$-dependence for $\ell\approx \ell_c$ is universal. 
If one uses a related observable, the average of the characteristic polynomial of the WL, a derived quantity from it is very well described by 
Burgers' equation with viscosity $\frac{1}{2N}$~\cite{Neuberger:2008mk}. 

Because perturbation theory amounts to path-integral integration  over the Lie algebra $su(N)$ and not over the group $SU(N)$ the large $N$ transition
at $\ell_c$ demarcates a scale where the prediction from perturbation theory depart substantially from the true values. That there exists an $\ell_c$~\cite{Blaizot:2008nc} does not imply confinement, only removes an obstacle to it at $\ell >\ell_c$, by closing the gap in the eigenvalue distribution. Confinement is reflected by the eigenvalue distribution approaching uniformity exponentially in 
$\ell^2$ as $\ell\to\infty$. This behavior occurs far from $\ell_c$ and has nothing to do with the mechanism that produced $\ell_c$ in the first place. 

Consider a theory that is expected to be scale invariant in the IR.
Let us consider the lattice version of single-angle eigenvalue distribution, $\rho(\theta,f;b, L)$, where $L$ is the linear extent of the loop in lattice
units. If $L < L_s(b)$, we will find a strong dependence of the distribution on $b$ and $L$ separately. But, we will be able to absorb it by finding a function $L(b)$ such that the distribution
only depends on $L(b)$. The resulting continuum distribution will exhibit the UV nature of the theory. 

If $L > L_l(b)$, we should expect the distribution
to stabilize and essentially become independent of $L$ and $b$. The resulting stable continuum distribution exhibits the scale invariance in the IR.
The distribution in the cross-over region, $ L_s(b) < L < L_l(b)$ will show dependence on $L$ and $b$. 
If we define {\it a lattice $\beta$-function}
with argument $\lambda=1/b$ in the crossover range by $[ L\frac{\partial}{\partial L} -\beta(\lambda)\frac{\partial}{\partial \lambda} ]\rho(\theta; \lambda, L)=0$
we shall find that $\beta(\lambda)$ comes out strongly $\theta$ dependent and is therefore not universal in any sense. These 
$\beta(\lambda)$ functions have lost all the power that the 
$\beta(\lambda)$ functions have at short distances, where they
are universal. 

A natural way to single out theories exhibiting ``perturbative IR behavior'' is to require that the lattice $\rho(\theta; L)$ have a gapped contiguous support on the unit circle $\forall L$, including at $L\to\infty$. This definition is at $N=\infty$.
If we have ``perturbative IR behavior'' there is reason to hope 
that a {\it simple} single coupling lattice approximation to the continuum RG trajectory holds. 

\bef
\begin{center}
\includegraphics[scale=0.6]{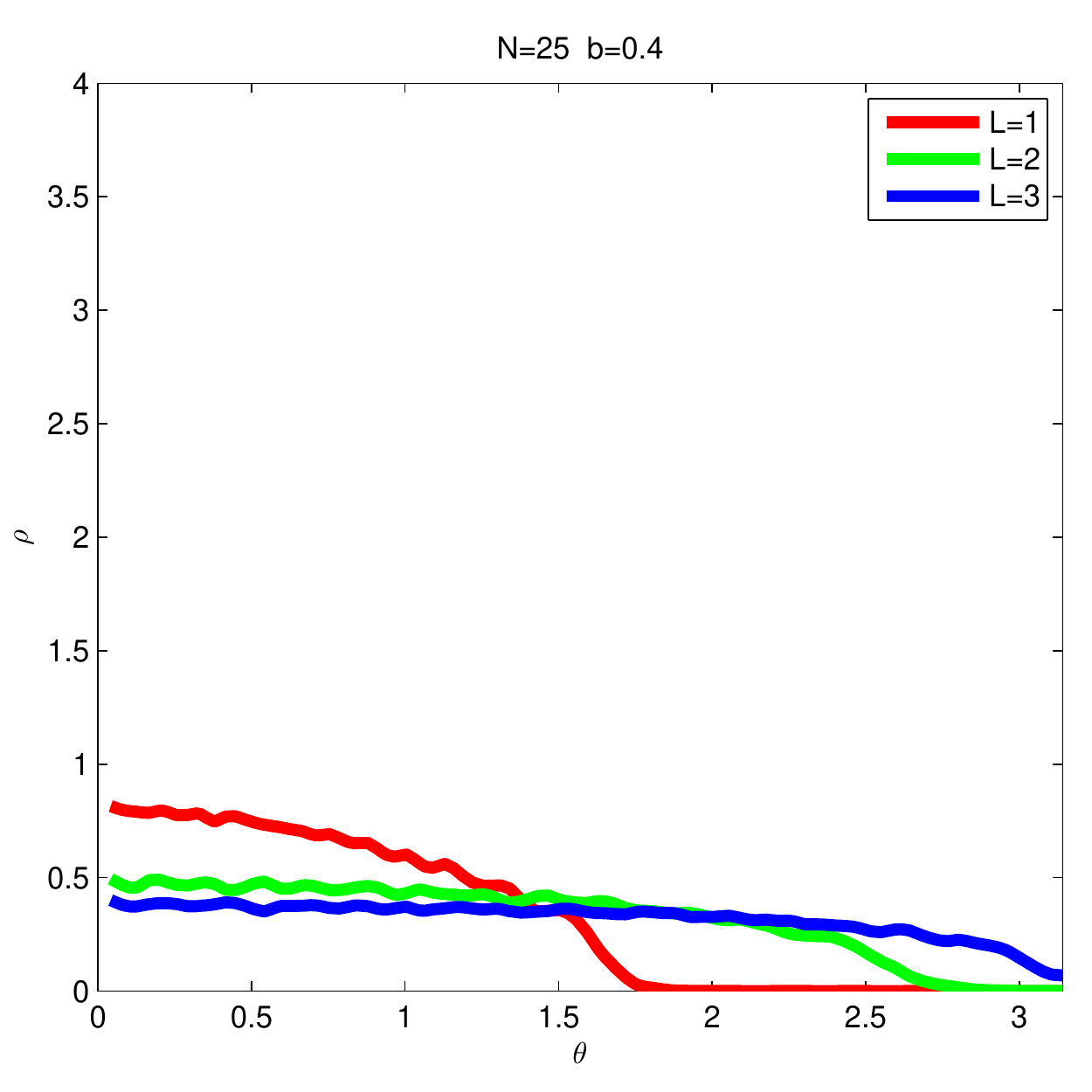}
\includegraphics[scale=0.6]{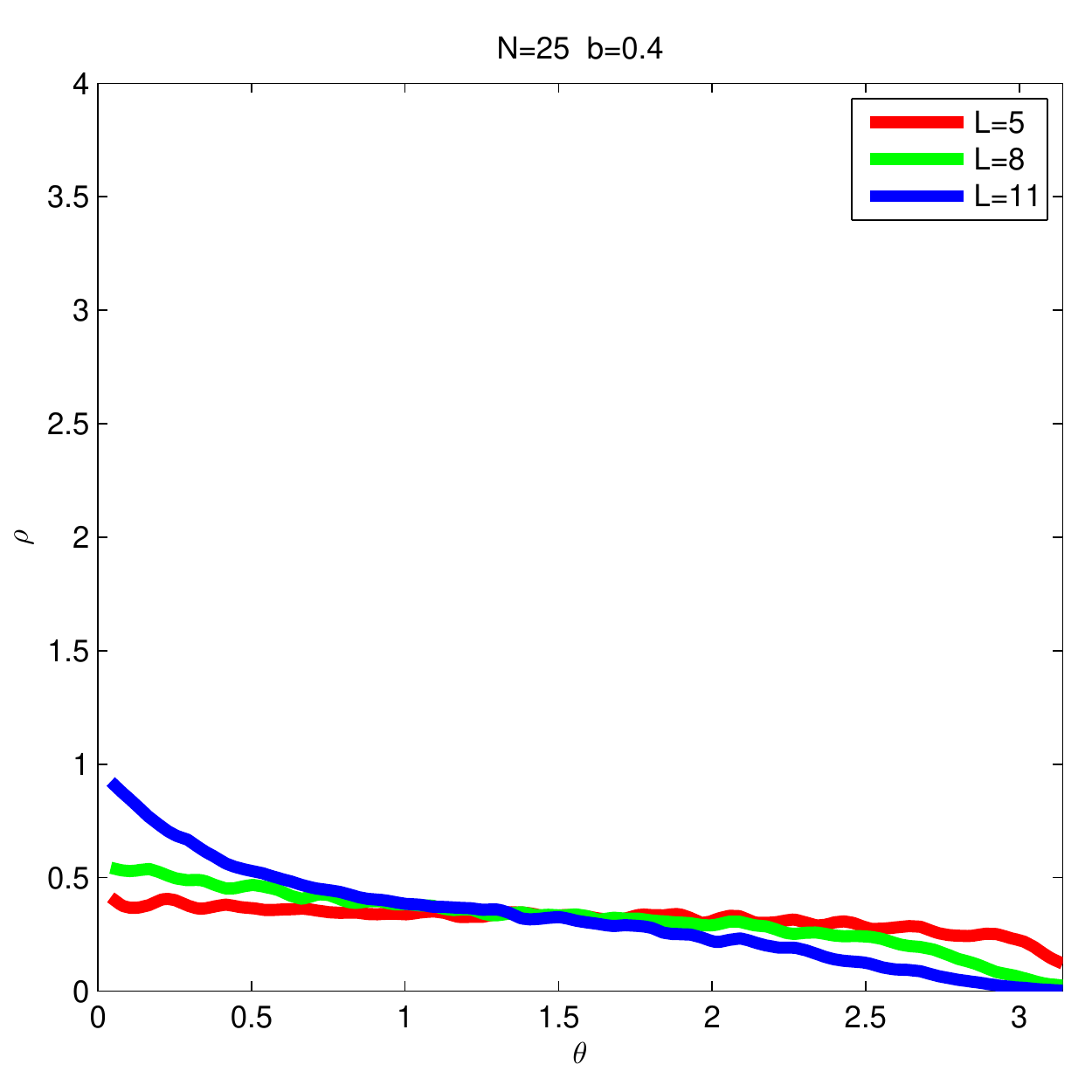}
\includegraphics[scale=0.6]{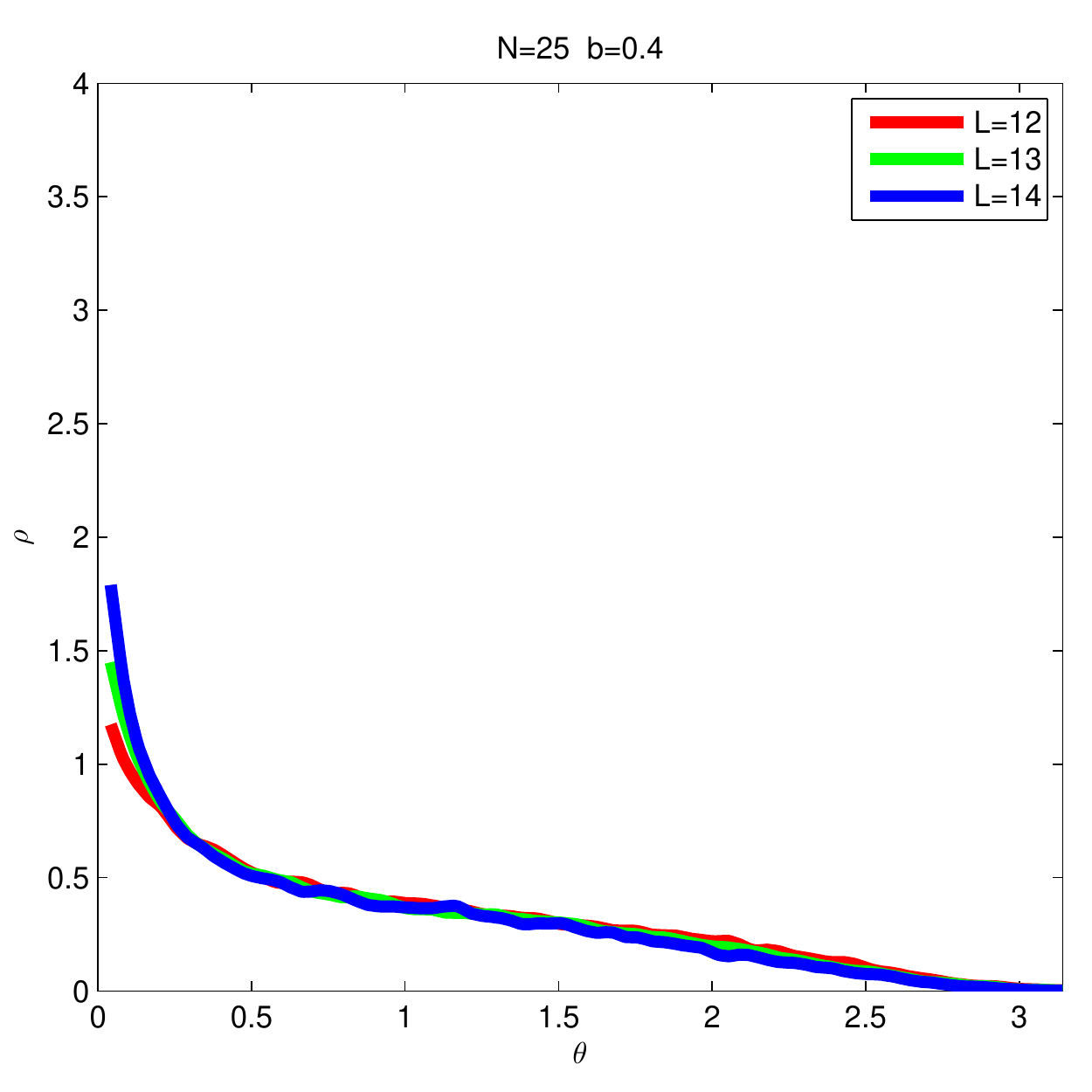}
\includegraphics[scale=0.6]{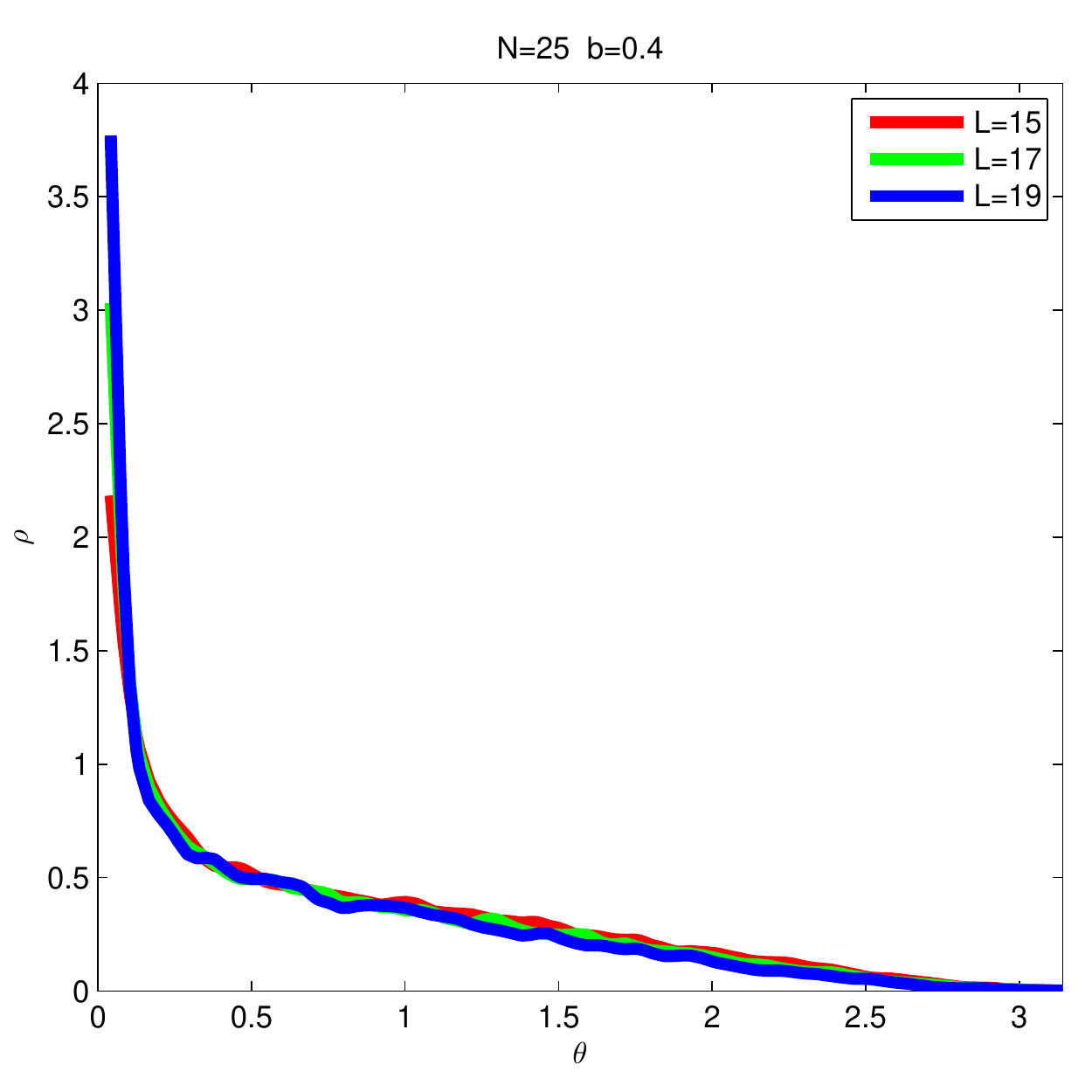}
\end{center}
\caption{
The distributions, $\rho(\theta,f;b,L)$, are shown for several different values of $L$ at $b=0.4$. The value of $N$ is $25$ and
the value of $f$ is set to $10^{-3}$. The value of $L$ increases as one goes from top-left to bottom-right with different panels
showing different regions of $L$.
}
\eef{fig1}

We will now provide numerical results for $\rho(\theta,f;b,L)$ in a matrix model where the IR behavior is not expected to be perturbative. The model is a single site SU(N) gauge theory 
coupled to two massless Majorana fermions realized using overlap fermions~\cite{Hietanen:2012ma}. We will provide results for $N=25$ which we consider 
to be large in the sense that the large $N$ limit for this particular lattice model is reasonably well approximated for most of the observables we look at. We will show data for three different values of lattice coupling, $b=0.4, 0.5, 0.6$. The lattice version of \eqn{smearing} is realized as follows:
Let $U_\mu(s) = e^{iA_\mu(s)}$ be the four smeared link variables on the single site lattice.
The smeared plaquettes are given by
\be
P_{\mu\nu}(s)= U_\mu(s) U_\nu(s) U_\mu^\dagger(s) U_\nu^\dagger(s).
\ee
Then,
\be
D_\nu F_{\mu\nu} (s) = 
i \sum_{\nu} \left[  P_{\mu\nu}^\dagger(s) + U_\nu^\dagger(s) P_{\mu\nu}(s) U_\nu(s) - P_{\mu\nu}(s) - U_\nu^\dagger(s) P_{\mu\nu}^\dagger(s) U_\nu (s)\right].
\ee
The evolution of $U_\mu(s)$ as per \eqn{smearing} becomes
\be
\frac{d U_\mu(s)}{ds} = -i D_\nu F_{\mu\nu}(s) U_\mu(s).
\ee
The folded and smeared Wilson loops used in the computation of $\rho(\theta,f;b,L)$ are
\be
W_{\mu\nu}(L,s=fL^2) = \left[U_\mu(s)\right]^L \left[ U_\nu(s)\right]^L \left[ U_\mu^\dagger(s)\right]^L \left [ U_\nu^\dagger(s)\right]^L.
\ee
\fgn{fig1} shows the results for $\rho(\theta,f;b,L)$ for different values of $L$ at $b=0.4$ and $f=10^{-3}$. The top-left panel shows the behavior
for $L< L_s(b)$. The $L=1$ and $L=2$ distributions show a gap with a larger gap for $L=1$. The $L=3$ distribution is gap-less.
The top-right panel shows the transition of $L$ into the region, $L_s(b) < L < L_l(b)$. The distributions still remain gap-less but increasing
$L$ results in a distribution that is more peaked around $\theta=0$. The bottom-left panel shows the development of scale invariant
distribution as $L$ moves closer to $L_l(b)$. The bottom-right panel shows distributions for $L > L_l(b)$ that are essentially
scale invariant. The distributions show a fast rise around $\theta=0$ but have stabilized away from $\theta=0$. This could be
indicative of a limiting distribution that has an integrable singularity at $\theta=0$. \fgn{fig2} shows the distribution as a function
of $b$ for $L=19$ which is above $L_l(b)$ for all three values of $b$. A scale invariant distribution seems to have emerged.

\bef
\begin{center}
\includegraphics[scale=1.0]{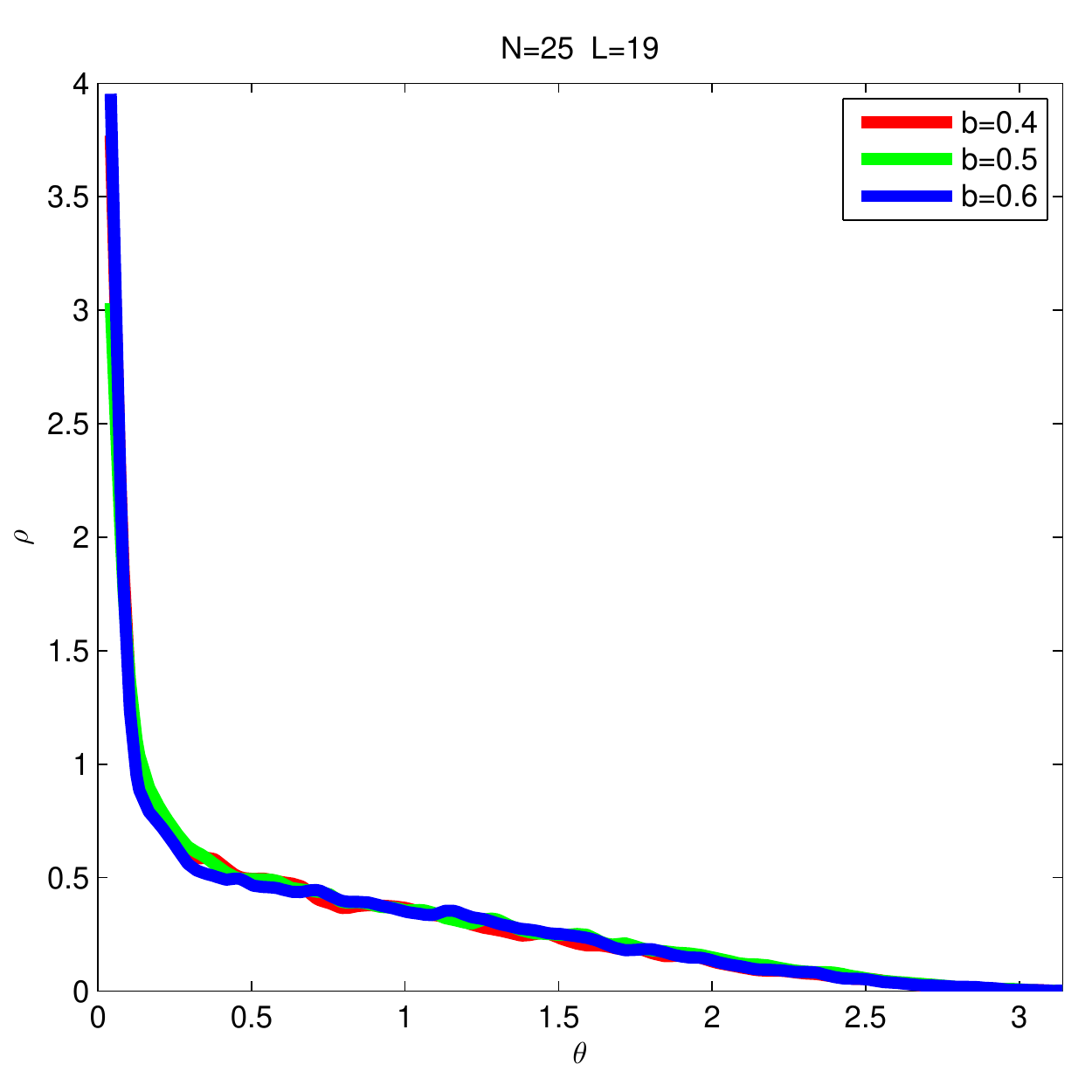}
\end{center}
\caption{
The distributions, $\rho(\theta,f;b,L)$, are shown for $b=0.4, 0.5, 0.6$ at $L=19$. The value of $N$ is $25$ and
the value of $f$ is set to $10^{-3}$. 
}
\eef{fig2}

We have demonstrated that scale invariance can emerge in a matrix model, indicating the existence of an IR FP.
The scales $L_s(b)$ and $L_l(b)$ have a ratio of order $5$ and
the set of lattice model for all couplings is far from making up an acceptable RG trajectory at intermediary scales.
The scale invariant angle distribution is gap-less. Therefore there is no indication for a concrete perturbative lattice realization of an IR FP.
We have demonstrated that
$\rho(\theta,f;b,L)$ is a useful probe of the theory. This probe  can be defined, and we suggest it will be useful, also at finite $N$.
 
This eigenvalue distribution reflects relatively simple physics for adjoint matter models: the force between fundamental test charges at distance $r$ is stronger than a Coulomb $1/r^2$ at short distances, 
raises to some constant value as $r$ increases further, but eventually turns around and goes like ${\rm Const}/r^2$ as $r\to\infty$. For a perturbative IR FP, the intermediary 
constant force regime would have to be skipped. A priori, there  seems to
be nothing prohibiting this from occurring, but this did not happen in our model. 

The matrix model of~\cite{Hietanen:2012ma} has the Wilson mass parameter that appears in the massless overlap fermion kernel set to $4$ in order to ensure proper 
reduction. In the present work we have re-analyzed data generated for ~\cite{Hietanen:2012ma} for lattice gauge couplings $b$ in the range $[0.32,0.70]$. Already in \cite{Hietanen:2012ma} lattice beta functions  were found to be far away from two
loop perturbation theory for this range of couplings.

We plan to continue our work with more extensive numerical studies on the lines of the present paper, using the
results of ~\cite{Lohmayer:2013spa} and ensuring that the center symmetries remain intact.
The formulation adopted here allows us to extend the matrix model to nonintergal numbers of massless flavors, a device that can, at the formal level, reduce the contribution of higher than two-loop terms to perturbative beta-functions at will. Also, it would increase our confidence in the relevance of the matrix model to the full lattice theory if we could increase $N$ relative to the $L$ values we now know we need. Recent interesting work on reduced model manages to deal with substantially larger values of $N$~\cite{Perez:2015yna}. This work uses also the device of twisted reduction. For an amount of matter $n_M\ge 1$ twisting, or any other trick on top of original Eguchi-Kawai reduction is not perturbatively required. Dropping twisting would simplify numerical work and might reduce the cost of simulations at $N$ of order 100 to something manageable on modest PC clusters. 

\acknowledgments

RN acknowledges partial support by the NSF under grant number PHY-1205396 and PHY-1515446. The research of HN was supported in part by the NSF under award PHY-1415525.

\bibliography{biblio}

\begin{thebibliography}{18}
\expandafter\ifx\csname natexlab\endcsname\relax\def\natexlab#1{#1}\fi
\expandafter\ifx\csname bibnamefont\endcsname\relax
  \def\bibnamefont#1{#1}\fi
\expandafter\ifx\csname bibfnamefont\endcsname\relax
  \def\bibfnamefont#1{#1}\fi
\expandafter\ifx\csname citenamefont\endcsname\relax
  \def\citenamefont#1{#1}\fi
\expandafter\ifx\csname url\endcsname\relax
  \def\url#1{\texttt{#1}}\fi
\expandafter\ifx\csname urlprefix\endcsname\relax\def\urlprefix{URL }\fi
\providecommand{\bibinfo}[2]{#2}
\providecommand{\eprint}[2][]{\url{#2}}

\bibitem[{\citenamefont{Sannino}(2009)}]{Sannino:2009za}
\bibinfo{author}{\bibfnamefont{F.}~\bibnamefont{Sannino}},
  \bibinfo{journal}{Acta Phys. Polon.} \textbf{\bibinfo{volume}{B40}},
  \bibinfo{pages}{3533} (\bibinfo{year}{2009}), \eprint{0911.0931}.

\bibitem[{\citenamefont{Lucini}(2015)}]{Lucini:2015noa}
\bibinfo{author}{\bibfnamefont{B.}~\bibnamefont{Lucini}}, \bibinfo{journal}{J.
  Phys. Conf. Ser.} \textbf{\bibinfo{volume}{631}}, \bibinfo{pages}{012065}
  (\bibinfo{year}{2015}), \eprint{1503.00371}.

\bibitem[{\citenamefont{Caswell}(1974)}]{Caswell:1974gg}
\bibinfo{author}{\bibfnamefont{W.~E.} \bibnamefont{Caswell}},
  \bibinfo{journal}{Phys. Rev. Lett.} \textbf{\bibinfo{volume}{33}},
  \bibinfo{pages}{244} (\bibinfo{year}{1974}).

\bibitem[{\citenamefont{Eguchi and Kawai}(1982)}]{Eguchi:1982nm}
\bibinfo{author}{\bibfnamefont{T.}~\bibnamefont{Eguchi}} \bibnamefont{and}
  \bibinfo{author}{\bibfnamefont{H.}~\bibnamefont{Kawai}},
  \bibinfo{journal}{Phys.Rev.Lett.} \textbf{\bibinfo{volume}{48}},
  \bibinfo{pages}{1063} (\bibinfo{year}{1982}).

\bibitem[{\citenamefont{Gonzalez-Arroyo and
  Okawa}(1983)}]{GonzalezArroyo:1982hz}
\bibinfo{author}{\bibfnamefont{A.}~\bibnamefont{Gonzalez-Arroyo}}
  \bibnamefont{and} \bibinfo{author}{\bibfnamefont{M.}~\bibnamefont{Okawa}},
  \bibinfo{journal}{Phys. Rev.} \textbf{\bibinfo{volume}{D27}},
  \bibinfo{pages}{2397} (\bibinfo{year}{1983}).

\bibitem[{\citenamefont{Bhanot et~al.}(1982)\citenamefont{Bhanot, Heller, and
  Neuberger}}]{Bhanot:1982sh}
\bibinfo{author}{\bibfnamefont{G.}~\bibnamefont{Bhanot}},
  \bibinfo{author}{\bibfnamefont{U.~M.} \bibnamefont{Heller}},
  \bibnamefont{and}
  \bibinfo{author}{\bibfnamefont{H.}~\bibnamefont{Neuberger}},
  \bibinfo{journal}{Phys. Lett.} \textbf{\bibinfo{volume}{B113}},
  \bibinfo{pages}{47} (\bibinfo{year}{1982}).

\bibitem[{\citenamefont{Mkrtchyan and Khokhlachev}(1983)}]{mkrtchyan:1982mk}
\bibinfo{author}{\bibfnamefont{R.}~\bibnamefont{Mkrtchyan}} \bibnamefont{and}
  \bibinfo{author}{\bibfnamefont{S.}~\bibnamefont{Khokhlachev}},
  \bibinfo{journal}{Pis'ma Zh. Eksp. Teor. Fiz.} \textbf{\bibinfo{volume}{37}},
  \bibinfo{pages}{160} (\bibinfo{year}{1983}).

\bibitem[{\citenamefont{Kovtun et~al.}(2007)\citenamefont{Kovtun, Unsal, and
  Yaffe}}]{Kovtun:2007py}
\bibinfo{author}{\bibfnamefont{P.}~\bibnamefont{Kovtun}},
  \bibinfo{author}{\bibfnamefont{M.}~\bibnamefont{Unsal}}, \bibnamefont{and}
  \bibinfo{author}{\bibfnamefont{L.~G.} \bibnamefont{Yaffe}},
  \bibinfo{journal}{JHEP} \textbf{\bibinfo{volume}{06}}, \bibinfo{pages}{019}
  (\bibinfo{year}{2007}), \eprint{hep-th/0702021}.

\bibitem[{\citenamefont{Hietanen and Narayanan}(2010)}]{Hietanen:2009ex}
\bibinfo{author}{\bibfnamefont{A.}~\bibnamefont{Hietanen}} \bibnamefont{and}
  \bibinfo{author}{\bibfnamefont{R.}~\bibnamefont{Narayanan}},
  \bibinfo{journal}{JHEP} \textbf{\bibinfo{volume}{01}}, \bibinfo{pages}{079}
  (\bibinfo{year}{2010}), \eprint{0911.2449}.

\bibitem[{\citenamefont{Narayanan and Neuberger}(2006)}]{Narayanan:2006rf}
\bibinfo{author}{\bibfnamefont{R.}~\bibnamefont{Narayanan}} \bibnamefont{and}
  \bibinfo{author}{\bibfnamefont{H.}~\bibnamefont{Neuberger}},
  \bibinfo{journal}{JHEP} \textbf{\bibinfo{volume}{03}}, \bibinfo{pages}{064}
  (\bibinfo{year}{2006}), \eprint{hep-th/0601210}.

\bibitem[{\citenamefont{Lohmayer and Neuberger}(2011)}]{Lohmayer:2011si}
\bibinfo{author}{\bibfnamefont{R.}~\bibnamefont{Lohmayer}} \bibnamefont{and}
  \bibinfo{author}{\bibfnamefont{H.}~\bibnamefont{Neuberger}},
  \bibinfo{journal}{PoS} \textbf{\bibinfo{volume}{LATTICE2011}},
  \bibinfo{pages}{249} (\bibinfo{year}{2011}), \eprint{1110.3522}.

\bibitem[{\citenamefont{Narayanan and Neuberger}(2007)}]{Narayanan:2007dv}
\bibinfo{author}{\bibfnamefont{R.}~\bibnamefont{Narayanan}} \bibnamefont{and}
  \bibinfo{author}{\bibfnamefont{H.}~\bibnamefont{Neuberger}},
  \bibinfo{journal}{JHEP} \textbf{\bibinfo{volume}{0712}}, \bibinfo{pages}{066}
  (\bibinfo{year}{2007}), \eprint{0711.4551}.

\bibitem[{\citenamefont{Lohmayer and Neuberger}(2012)}]{Lohmayer:2011nq}
\bibinfo{author}{\bibfnamefont{R.}~\bibnamefont{Lohmayer}} \bibnamefont{and}
  \bibinfo{author}{\bibfnamefont{H.}~\bibnamefont{Neuberger}},
  \bibinfo{journal}{Phys. Rev. Lett.} \textbf{\bibinfo{volume}{108}},
  \bibinfo{pages}{061602} (\bibinfo{year}{2012}), \eprint{1109.6683}.

\bibitem[{\citenamefont{Neuberger}(2008)}]{Neuberger:2008mk}
\bibinfo{author}{\bibfnamefont{H.}~\bibnamefont{Neuberger}},
  \bibinfo{journal}{Phys. Lett.} \textbf{\bibinfo{volume}{B666}},
  \bibinfo{pages}{106} (\bibinfo{year}{2008}), \eprint{0806.0149}.

\bibitem[{\citenamefont{Blaizot and Nowak}(2008)}]{Blaizot:2008nc}
\bibinfo{author}{\bibfnamefont{J.-P.} \bibnamefont{Blaizot}} \bibnamefont{and}
  \bibinfo{author}{\bibfnamefont{M.~A.} \bibnamefont{Nowak}},
  \bibinfo{journal}{Phys. Rev. Lett.} \textbf{\bibinfo{volume}{101}},
  \bibinfo{pages}{102001} (\bibinfo{year}{2008}), \eprint{0801.1859}.

\bibitem[{\citenamefont{Hietanen and Narayanan}(2012)}]{Hietanen:2012ma}
\bibinfo{author}{\bibfnamefont{A.}~\bibnamefont{Hietanen}} \bibnamefont{and}
  \bibinfo{author}{\bibfnamefont{R.}~\bibnamefont{Narayanan}},
  \bibinfo{journal}{Phys. Rev.} \textbf{\bibinfo{volume}{D86}},
  \bibinfo{pages}{085002} (\bibinfo{year}{2012}), \eprint{1204.0331}.

\bibitem[{\citenamefont{Lohmayer and Narayanan}(2013)}]{Lohmayer:2013spa}
\bibinfo{author}{\bibfnamefont{R.}~\bibnamefont{Lohmayer}} \bibnamefont{and}
  \bibinfo{author}{\bibfnamefont{R.}~\bibnamefont{Narayanan}},
  \bibinfo{journal}{Phys. Rev.} \textbf{\bibinfo{volume}{D87}},
  \bibinfo{pages}{125024} (\bibinfo{year}{2013}), \eprint{1305.1279}.

\bibitem[{\citenamefont{Garc\'ia~P\'erez
  et~al.}(2015)\citenamefont{Garc\'ia~P\'erez, Gonz\'alez-Arroyo, Keegan, and
  Okawa}}]{Perez:2015yna}
\bibinfo{author}{\bibfnamefont{M.}~\bibnamefont{Garc\'ia~P\'erez}},
  \bibinfo{author}{\bibfnamefont{A.}~\bibnamefont{Gonz\'alez-Arroyo}},
  \bibinfo{author}{\bibfnamefont{L.}~\bibnamefont{Keegan}}, \bibnamefont{and}
  \bibinfo{author}{\bibfnamefont{M.}~\bibnamefont{Okawa}},
  \bibinfo{journal}{JHEP} \textbf{\bibinfo{volume}{08}}, \bibinfo{pages}{034}
  (\bibinfo{year}{2015}), \eprint{1506.06536}.

\end{thebibliography}
\end{document}